\documentclass[twocolumn,
aps,nofootinbib,showpacs,showkeys,preprint
tightenlines,preprintnumbers,superscriptaddress]{revtex4}

\usepackage{epsf,epsfig,subfigure,graphicx,amsmath,amssymb}
\usepackage{color}

\newcommand{\dis}[1]{\begin{equation}\begin{split}#1\end{split}\end{equation}}

\newcommand{\ie}{{\it i.e.}\ }
\newcommand{\etal}{{\it et al.} }

\begin{document}

\title{\bf Calculations of resonance enhancement factor in axion-search tube-experiments}

\author{Jooyoo Hong}
\address{Department of Applied Physics, Hanyang University at Ansan, Ansan, Kyonggi-Do 426-791, Korea,}
\author{Jihn E. Kim}
\address{Department of Physics, Kyung Hee University, 26 Gyeongheedaero, Dongdaemun-Gu, Seoul 130-701, Korea,}

\author{Soonkeon Nam}
\address{Department of Physics, Kyung Hee University, 26 Gyeongheedaero, Dongdaemun-Gu, Seoul 130-701, Korea,}

\author{Yannis Semertzidis}
\address{Center for Axion and Precision Physics Research, Institute for Basic Science(IBS),
 291 Daehakro, Yuseong-Gu, Daejeon 305-701, Republic of Korea,}
\address{Depatyment of Physics, KAIST, Daejon 305-701, Republic of Korea}

\begin{abstract}

It is pointed out that oscillating current density, produced due to the coupling between an external magnetic field and the cosmic axion field, can excite the TM resonant modes inside an open-ended cavity (tube). By systematically solving the field equations of axion-electrodynamics we obtain explicit expressions for the oscillating fields induced inside a cylindrical tube. We calculate the enhancement factor when a resonance condition is met. While the power obtained for TM modes replicates the previous result, we emphasize that the knowledge of explicit field configurations inside a tube opens up new ways to design axion experiments including a recent proposal to detect the induced fields using a superconducting LC circuit. In addition, as an example, we estimate the induced fields in a cylindrical tube in the presence of a static uniform magnetic field applied only to a part of its volume.

\end{abstract}


\maketitle


{\it Introduction}--
The discovery of the axion would be a very exciting event in itself as well as in the context of cold dark matter (CDM) searches. It would be a culmination of various interesting theoretical ideas sparked by the discovery of $\theta$  vacuum  \cite{QCDangle} of quantum chromodynamics\,(QCD) in the  late 1970s \cite{Revs}. The original Peccei-Quinn-Weinberg-Wilczek(PQWW) axion \cite{PQ77, PQWW} has been excluded in early experiments, but the axion has been resurrected under the name of `invisible axion' \cite{KSVZ, DFSZ} which may be biding its time \cite{cosmicaxion} behind the curtain of the axion window \cite{Bae08}. It may be possible to detect this very light, weakly interacting particle dubbed `invisible axion'  in the cavity-type detectors \cite{sikivie83,sikivie85,Krauss85}.

A smoking gun proof of axions in the cosmic context  is the detection of the oscillating nature of the axion field. The oscillating current induced by coherent motion of  axions is a novel feature whose detection possibility was first suggested a long time ago \cite{Hong90}. In recent years, a few proposals to detect the oscillating nature of CDM axion have been made \cite{graham,sikivie13}, igniting a new interest in this direction.
In particular an experiment proposed in Ref. \cite{sikivie13} employing a superconducting LC circuit to pick up signals from oscillating magnetic flux is conceptually simple and is on the threshold of technical feasibility.  In this Letter, we examine the possibility to enhance the induced signals by surrounding the solenoidal region with a metallic wall of high conductivity, thus capitalizing on resonant enhancement of cavity modes. We find that the current signal to be detected by this LC circuit detector can get a significant boost from the large resonance enhancement factor  once the resonance condition is met. Though this resonance enhancement itself is nothing new and was envisioned a long time ago in the context of axion searches\,\cite{Krauss85}, explicit mode calculation has not been provided, whereas we argue here that the knowledge of explicit field configuration inside a metallic tube is essential for the optimal positioning of the pickup loop to squeeze out the strongest output signals.

For a tube of cylindrical geometry we present a complete derivation of the resonant modes and the associated physical quantities excited by coherent oscillation of relic axions coupled with the static magnetic field.

The interaction between axion and electromagnetic fields modifies Maxwell's equations, which at very low energies can be described by the following effective Lagrangian
\dis{
{\cal L} =& \frac{1}{2} \partial_\mu a~ \partial^\mu a - V(a)
 + \frac{1}{4}g a F_{\mu\nu} {\tilde F}^{\mu\nu} \\
 &- \frac{1}{4} F_{\mu\nu} F^{\mu\nu} +j^\mu A_\mu + a \rho_a ,\label{eq:axemL}
}
where $F_{\mu\nu}$ is the electromagnetic field strength ({\bf E, B}) of electromagnetic potential $A_\mu$,  $\tilde{F}_{\mu\nu}$ its dual $\tilde{F}_{\mu\nu}=\frac12\epsilon_{\mu\nu\rho\sigma} F^{\rho\sigma}$, and $j_\mu$ the electromagnetic current density. The axion potential $V(a)$ will be approximated as the quadratic axion mass term, $(1/2)m_a^2a^2$, which would be sufficient for our purpose of calculating the effect of coherent oscillation of  axions.  $\rho_a\equiv g_5\rho_5$ depicts the pseudoscalar axionic charge density. For electrons, for instance, the coupling is $g_5=g_e \frac{m_e}{f_a}$ when $\rho_5=\bar{\psi}_e i\gamma_5\psi_e$ represents the axial charge density of electron. The axion-photon coupling constant $g$ is a combination of a model-dependent axion coupling parameter $g_\gamma$ \cite{Kim98} and the axion decay constant $f_a$ as well as the fine structure constant:  $g=g_\gamma \frac{\alpha_{em}}{\pi f_a}$.


{\it Axion-electrodynamics for tube resonator}-- In terms of {\bf E} and {\bf B}, the field equations are given by
\dis{
& {\nabla} \cdot {\bf E} = \rho + g {\nabla}a \cdot {\bf B} , \\[0.3em]
& {\nabla} \times{\bf B}  - \partial_t {\bf E}  = {\bf j}
   - g{\bf B} \partial_t a  - g{\nabla}a \times
   {\bf E} ,  \\[0.3em]
& {\nabla} \cdot{\bf B}  = 0, \\[0.3em]
& {\nabla} \times {\bf E} + \partial_t{\bf B}  = 0,\\[0.3em]
& ( \partial_t^2 - {\nabla}^2 )a
        =-V'(a) - g {\bf E}  \cdot {\bf B} + \rho_a.  \label{axionem}
}
In the last equation, the use of a quadratic axion potential results in $V'(a)\simeq m_a^2 a$.

The cosmic axion or halo axion field is homogeneous over a large de Broglie wavelength and oscillates about the potential minimum in a coherent way, which renders it a good candidate for DM. The axion oscillation $a(t)$ contributes to the cosmic energy density $m_a^2\langle a^2(t)\rangle$ denoted as  $\rho_D$ here which should not overclose the energy density of the Universe. Thus, the axion mass and the initial vacuum misalignment angle $\theta_i=a/f_a$ is constrained from the observed value of $\rho_D$ \cite{Bae08}.

Let us solve Eq. (\ref{axionem}) order by order for a right cylindrical tube of radius $R_c$, length $\ell$ and conductivity $\sigma$  inside the volume of which an external static magnetic field is applied longitudinally by a solenoidal coil of current $I$ and turns per unit length $n$. At the lowest order, from the charge and current densities ($\rho_0=0,\ {\bf j}_0=nI\delta(r-R_c)\hat{\theta}$),  we have ${\bf E}_0=0,  {\bf B}_0=nI\,\theta(R_c-r)\hat{z}$ where $\theta(x)$ is the Heaviside step function.

Since the coherence of the halo axion oscillation is spoiled by a small amount due to thermalization process in the Milky Way, the axion field will be represented by the following Fourier expansion ($T$ is chosen as the long time over which the average will be taken)
\dis{
a(t) &=\sqrt{T}\int_{-\infty}^{+\infty} \frac{d\omega}{2\pi} {\cal A}(\omega) e^{-i\omega t}\\
{\cal A}(\omega) &=\frac{1}{\sqrt{T}}\int_{-T/2}^{T/2}\, dt\, a(t) e^{i\omega t}.
}
The average of the axion field squared satisfies the Parseval relation,
\dis{
\langle a^2(t)\rangle =\frac{1}{T} \int_{-T/2}^{T/2}\, dt\, a^2(t)  = \int_{-\infty}^{+\infty} \frac{d\omega}{2\pi} |{\cal A}(\omega)|^2.
}
Assuming the axion halo conforms to the Maxwellian momentum distribution which is sharply peaked around $\omega=m_a(1+\langle v^2\rangle/6)\equiv \omega_a$ \cite{Krauss85}, the power spectrum of oscillation can be faithfully mimicked by the Breit-Wigner function,
\dis{
|{\cal A} (\omega)|^2=\frac{\omega_a \rho_{D}}{m_a^2 Q_a}\, \frac{1}{(\omega-\omega_a)^2 +(\omega_a/2Q_a)^2}
}
where $Q_a$ is defined as the width of axion velocity distribution
\dis{
Q_a\equiv \frac{\omega_a}{\Delta \omega}\simeq \frac{m_a}{m_a \langle v^2\rangle/3}\sim 3\times 10^{6}.
}

At the next order, we have relations between small induced fields and currents which would constitute a TM mode.
Both inside the tube and within the conductor, the  electric field is longitudinal ${\bf E}_1=E_z(r,t)\hat{z}$ and the magnetic field is azimuthal ${\bf B}_1=B_\theta(r,t) \hat{\theta}$.
Again, let us decompose these fields into the Fourier components
\dis{
B_\theta(r,t) &=\sqrt{T}\int_{-\infty}^{+\infty}\frac{d\omega}{2\pi}\,B(r,\omega)\,e^{-i\omega t},  \\[0.3em]
E_z(r,t) &=\sqrt{T}\int_{-\infty}^{+\infty}\frac{d\omega}{2\pi}\,E(r,\omega)\,e^{-i\omega t}.
}
\begin{figure}[!t]
  \begin{center}
  \begin{tabular}{c}
   \includegraphics[width=0.15\textwidth]{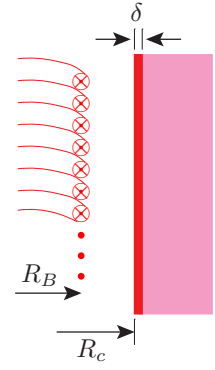}
  \end{tabular}
  \end{center}
 \caption{The skin depth $\delta$ drawn from the tube wall. The cylindrical tube has radius $R_c$ and length $\ell$. The static external magnetic field ${\bf B}_0$ is applied longitudinally for the cylindrical volume of radius $R_B\leq R_c$.
  }
\label{fig:BC}
\end{figure}
Then, inside the tube the equations in Eq. (\ref{axionem})  become
\dis{
\begin{array}{l}
  B'  +\frac{1}{r}B+i\omega E   =  i\omega gnI{\cal A}(\omega), \\ [0.3em]
  - E' -i\omega B = 0, \end{array}
}
from which we obtain
\dis{
&E(r,\omega) = gnI{\cal A}(\omega) +CJ_0(\omega r)  ,\\
&B(r,\omega)=-iC J_1(\omega r)\, \,
}
where $C$ is to be determined from the conditions at the  inner wall of the tube. In the conductor ($r>R_c$)  we have the following equations
\dis{
\begin{array}{l}
B'  +\frac{1}{r}B +i\omega E  = \sigma E , \\ [0.3em]
  -E'-i\omega  B = 0,
  \end{array}
}
where the current density $j_1=\sigma E_1\theta(r-R_c)$ is nonzero inside the conductor and is given by Ohm's law with conductivity $\sigma$. (Refer to Fig. \ref{fig:BC} in the limit of $R_B\to R_c$.) For a good conductor, it is expected that fields would decrease very sharply in $r$, giving an effective skin depth $\delta$. Thus, we use the approximation $ B' \gg B /r$ and $\sigma\gg \omega\simeq m_a\,(\sigma\simeq 6\times 10^{17}\,\textrm{Hz for Cu},\, m_a<10^{12}\textrm{ Hz for cosmic axions})$. Therefore, the equations ($B'  \simeq\sigma E $ and $-E'-i\omega  B=0$) can be solved to give the magnetic field inside the  conductor
\dis{
B(r,\omega )=-iC J_1(\omega R_c)\, e^{-(1-i)(r-R_c)/\delta}
}
where the continuity of tangential component of {\bf B} at $r=R_c$ was used to fix the field at the wall ($r=R_c$). As usual the skin depth $\delta$ is defined as
$\delta=(2/\sigma \omega )^{1/2}$. Then in turn the magnetic field $B(r,\omega)$  determines the electric field $E(r,\omega)$ in the conductor as
\dis{
E(r,\omega )=\frac{i(1-i)}{\sigma\delta}\,C\,J_1(\omega  R_c) e^{-(1-i)(r-R_c)/\delta}.
}
Note that inside the imperfect conductor the electric field is nonzero within the skin depth though much smaller than  the magnetic field.

{\it Resonance enhancement factor}--
The continuity of $E $- field at the boundary $r=R_c$ gives a relation $gnI{\cal A}(\omega ) + C J_0(\omega R_c)= i(1-i)(C/\sigma\delta) J_1(\omega R_c)$, and determines $C$ as
\dis{
C=-gnI{\cal A}(\omega )\, Q_J(\omega )\, e^{-i\phi_J(\omega )}
}
where
\dis{
&Q_J = \left(J_0^2 -\sqrt{\frac{2\omega }{\sigma}} J_0  J_1 + {\frac{\omega }{\sigma}}J_1^2 \right)^{-1/2},\\
&\phi_J =-\tan^{-1}\frac{\sqrt{\omega /2\sigma}\, J_1}{J_0-\sqrt{\omega /2\sigma}\,J_1},
}
Here the arguments of $J_0$ and $J_1$ are $\omega R_c$. On resonances where $C$ becomes huge, $J_0(\omega R_c)=0$, \ie $ \omega  R_c=\chi_{0l}$ for the $l$-th zero of $J_0(x)$. At the frequencies where this condition is satisfied, we obtain a huge enhancement factor,
\dis{
Q_J =\sqrt{\frac{\sigma}{\omega }}\, \frac{1}{|J_1(\chi_{0l})|},\qquad
\phi_J  =l\pi +\frac{\pi}{4}.
}
Since the factor $Q_J(\omega )$ is substantial only at the resonances while it is of order 1 away from resonances, it is useful to expand the above formula near the resonance points, $\omega  R_c=\chi_{0l}$. Then we get the following expression which clearly shows the resonant behavior.
\dis{
Q_J(\omega )e^{-i\phi_J(\omega )}\simeq -\frac{1}{RJ_1(\chi_{0l})}\cdot \frac{1}{(\omega -\omega_l)+i\frac{\omega _l}{2Q_l}}
}
where $\omega _l$ and $Q_l$ are defined at the resonances, $\omega_l\simeq\chi_{0l}/R_c$ and $Q_l=\sqrt{\chi_{0l}\, \sigma R_c/2}$. When $\omega= \chi_{0l}/R_c$, $Q_l=R_c/\delta$ which is the quality factor for the tube wall.

\begin{figure}[!b]
  \begin{center}
  \begin{tabular}{c}
   \includegraphics[width=0.45\textwidth]{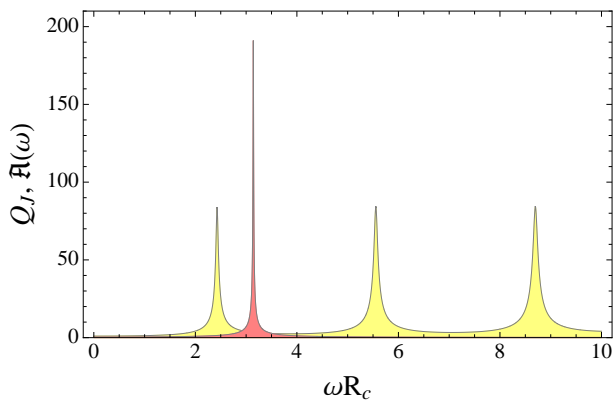}
  \end{tabular}
  \end{center}
 \caption{
 An illustration of axion-tube resonance($Q_l\sim 100$, $Q_a\sim 300$, $m_a\sim 1.3\,\omega_1$).  The yellow curve is $Q_J$ as a function of $\omega R_c$ and the reddish one is $|{\cal A}(\omega)|$ peaked at $m_aR_c$. A resonance occurs when the center of $|{\cal A}(\omega)|$ coincides with one of the peaks of $Q_J$, {\it i.e.} $\omega_l\simeq m_a$. The actual resonance curves would be much sharper.
  }
\label{fig:QJform}
\end{figure}


{\it Cavity TM mode}--
Once  the constant $C$ is obtained, we have the following expressions for the electromagnetic fields and current,
\begin{widetext}
\dis{\textrm{Cavity}\left\{
\begin{array}{l}
E^{\rm cav}_z(r,t)  =\sqrt{T} \int_{-\infty}^\infty\frac{d\omega }{2\pi}\,gnI{\cal A}(\omega )[1-J_0(\omega r)Q_J(\omega )e^{-i\phi_J(\omega )} ]e^{-i\omega t} ,\\[0.3em]
B^{\rm cav}_\theta(r,t) =\sqrt{T} \int_{-\infty}^\infty\frac{d\omega }{2\pi}\,ignI{\cal A}(\omega ) J_1(\omega r)Q_J(\omega )e^{-i\phi_J(\omega )}  e^{-i\omega t},
\end{array} \right.
}

\dis{\textrm{Conductor}\left\{
\begin{array}{l}
E^{\rm con}_z(r,t)  =\sqrt{T} \int_{-\infty}^\infty\frac{d\omega }{2\pi}\,\frac{-i(1-i)}{\sigma\delta}gnI{\cal A}(\omega ) J_1(\omega R_c)Q_J(\omega )e^{-i\phi_J(\omega )} e^{-i\omega t  -(1-i)(r-R_c)/\delta} ,\\[0.3em]
B^{\rm con}_\theta(r,t) =\sqrt{T} \int_{-\infty}^\infty\frac{d\omega }{2\pi}\,ignI{\cal A}(\omega ) J_1(\omega R_c)Q_J(\omega )e^{-i\phi_J(\omega )}  e^{-i\omega t -(1-i)(r-R_c)/\delta},    \end{array}\right.
 }
\end{widetext}
and $j_z(r,t)  =\sigma E^{\rm con}_z(r,t)$ for $r\geq R_c$.

{\it Energy  and power loss}--  Inside the tube the total stored energy is
\dis{
U &=\int_V\,d^3r \left\langle\frac{{\bf E}_1^2+{\bf B}_1^2}{2} \right\rangle \\
&= (gnI )^2 V \frac{\omega_a\rho_D}{ (m_aR_c)^2Q_a}\int_{-\infty}^\infty\frac{d\omega }{2\pi}\,F(\omega,\omega_a)F(\omega,\omega_l).
}
In deriving this result we assumed a good conductor. The integrand is the product of two Breit-Wigner functions
\dis{
F(\omega,\omega_{a,l})=\frac{1}{(\omega-\omega_{a,l})^2+(\frac{\omega_{a,l}}{2Q_{a,l}})^2 }.
}
The centers and widths of two Breit-Wigner functions are different in general. Unless the two centers almost coincide, i.e. without resonances, the total energy $U$ is insignificant.
On the other hand, a resonance occurs when the two centers coincide, i.e. $\omega_l\simeq \omega_a$  (see Fig. \ref{fig:QJform}), implying a large enhancement for the stored energy $U$. Depending on the relative magnitude of  $Q_a$ and $Q_l$,  on resonance we can obtain the total energy as
 \dis{
U  \simeq(gnI )^2 V\frac{\omega_a\rho_D}{ (m_aR_c)^2Q_a}\left\{\begin{array}{cc}\frac{4Q_aQ_l^2}{\omega_a^3}\quad &(Q_a\gg Q_l),\\ [0.5em] \frac{4Q_lQ_a^2}{\omega_a^3}\quad&(Q_a\ll Q_l).\end{array}\right.
}
The Ohmic power loss in the conductor can also be computed
\dis{
P& =\int d^3r \,\langle{\bf j}\cdot{\bf E}^{\rm con} \rangle \\
&=(gnI)^2 S \sqrt{\frac{\omega_a}{2\sigma}}\frac{\omega_a\rho_D}{\, (m_aR_c)^2Q_a}\left\{\begin{array}{cc}\frac{4Q_aQ_l^2}{\omega_a^3}\quad &(Q_a\gg Q_l),\\ [0.5em] \frac{4Q_lQ_a^2}{\omega_a^3}\quad&(Q_a\ll Q_l),\end{array}\right. \label{ploss}
}
where $S=2\pi R\ell$ is the tube wall area. We can check that the same result follows by calculating
the Poynting vector at the tube wall, $\langle{\bf E}_1\times {\bf B} _1 \rangle =\hat{r}\langle-E_z(R_c,t)B_\theta(R_c,t)\rangle $.

The quality factor for the tube can also be checked to be of the expected form
\dis{
Q_{\rm tube} =\omega_a \frac{\textrm{Stored energy} }{\textrm{Power loss} }=\omega_a\frac{U}{P }
=\frac{2V}{S\delta}=\frac{R_c}{\delta}=Q_l.
}
This gives $Q_{\rm tube} \sim 10^5$, for instance,  for a cylindrical tube of $R_c=0.2$\,m (with the resonance condition $m_aR_c=\chi_{01}\simeq 2.4$) made of copper and for axion mass of $m_a\sim 3.6$\,GHz. Compared to this tube $Q$-factor, for cosmic axions $Q_a\sim 3\times 10^6$.

The power absorbed by the tube wall in Eq. (\ref{ploss}) can be equivalently obtained as follows. Using $Q_{\rm tube}=Q_l$, we get
\dis{
P=\frac{\omega_aU}{Q_l}= \frac{(gB_0)^2\rho_{D}V}{m_a}\cdot\frac{4}{\chi_{0l}^2}\ {\rm min}(Q_l,Q_a)
}
where we use the notation for the external magnetic field $B_0=nI$ and the relation $\omega_a\simeq m_a$. Here we see that the result in Ref. \cite{sikivie85} is reproduced precisely.

{\it Solenoid enclosed in a tube resonator}--  The explicit calculation of resonant modes allows us to consider giving variations in the design of tubes or cavities for axion DM search. We can divide the volume of a tube into a region applied by a magnetic field and the rest free of the external magnetic field. To be more specific, we consider a cylindrical tube of radius $R_c$. We then have an external magnetic field $B_0=nI$ applied longitudinally over a region of a concentrical cylinder of radius $R_B\,(R_B<R_c)$, as shown in Fig. \ref{fig:BC}.

If the coils in the solenoid are wound in a single thread entering from one end exiting at the other, there would be a net longitundinal current $I$ which produces a  static azimuthal magnetic field ($(I/2\pi r)\hat{\theta}$) outside the solenoid, which can excite a TE mode. However, this azimuthal field is much weaker than the longitudinal magnetic field $B_0=nI$ with the ratio between them smaller than $1/2\pi nR_B$ which is practically negigible for a very large value of $n$. We can also make this azimuthal field even weaker by winding the coils in a way entering and exiting at the same end for cancellation of unwanted magnetic fields.  In a real experiment, however,  even a weak fringe magnetic field may wreak havoc when combined with noise in the current. In that case we need to shield out magnetic fields outside the solenoid to make our analysis more precise and definite.

Now assuming a static external magnetic field is nonzero only inside the solenoid, as previously done, we can proceed to solve equations in each region and match the boundary conditions to get the resonant modes for the tube. While in the region $r\leq R_B$ only the Bessel functions of the first kind are needed to write general solutions, in the region $R_B\leq r\leq R_c$ we need also the Bessel functions of the second kind (which are finite everywhere in this region) to conform to the boundary conditions both at $r=R_B$ and at $r=R_c$.
We find that when a resonance occurs a single coefficient (denoted as $C$ before) dominates over others and thus the resonant tube modes are essentially the same as the ones for a fully filled tube except for a multiplicative factor $f$ which is less than $1$,
\dis{
&E(r,\omega)\simeq f\ E_{\rm filled}(r,\omega)\\[0.3em]
&B(r,\omega)\simeq f\ B_{\rm filled}(r,\omega)
 }
where `filled' points to the case of a tube completely filled by ${\bf B}_0\,(\ie~R_B\to R_c)$ and the so-called {\it filling factor} $f$ is defined as
\dis{
f\equiv \frac{x_B J_1(x_B)}{\chi_{0l}J_1(\chi_{0l})}.
}
For $x_B\ll 1$, the filling factor becomes $f\simeq \frac{\chi_{0l}}{2J_1(\chi_{0l})}\frac{R_B^2}{R_c^2}$. In other words the field amplitude is reduced roughly by the ratio between the cross-sectional areas of the  solenoid and the tube, which makes sense.
The coefficient $C$ that multiplies the Bessel functions of $E_{\rm filled}$ and $B_{\rm filled}$ is given by
\dis{
C =&gnI{\cal A}(\omega)\frac{\pi x_B}{2}\Big( N_1(x_B)\\
&-J_1(x_B)\frac{N_0(x_c) -\frac{i(1-i)}{\sigma\delta}N_1(x_c)}{J_0(x_c)-\frac{i(1-i)}{\sigma\delta}J_1(x_c)} \Big)
}
where $x_B=\omega R_B$ and $x_c=\omega R_c$. In the limit $x_B\to x_c$, we recover $C$ obtained in the previous section.

This kind of alternative tube geometry can be quite advantageous when we try to directly measure the electromagnetic field inside a resonant tube by inserting a pickup loop. If the induced magnetic field reaches its maximum outside the magnetic field region, we can install a superconducting pickup loop without being hampered by an external magnetic field or any equipment to produce it.

And since we can make the size of tube bigger without worrying about supplying a strong magnetic field for the whole volume of the tube, it would be much easier to meet the resonance conditions for low-mass axions or ALPs, though we have to pay for the reduced strength of power by $f^2$. An alternative way of recovering the cavity quality factor might be using a microwave cavity filled by a lossless dielectric with large dielectric coefficient at cryogenic temperatures \cite{caspers}.

To make sure that we hit any of resonance conditions we should  change the radius of the tube $R_c$ in a continuous way for a wide range of its value. For example, we may make a tube wall by rolling a very flexible metal sheet into a tightly wound cylinder of some initial radius, say $R_B$. Then we loosen the clasp in a slow but controlled way to increase  the tube radius continuously while monitoring the signal to see if any resonance occurs. Technical details would not be discussed here, but we argue that with this effective tuning method it would be possible to scan the whole axion window in a time scale much shorter than conventional methods.

{\it An LC circuit experiment}--  As in the experimental setup proposed in Ref. \cite{sikivie13}, an LC circuit can be introduced which consists of a  rectangular pickup loop of length $\ell_p$ and an accompanying coaxial cable leading up to a detector coil. The pickup loop is parallel to the tube and spans from $r=R_1$ to $r=R_2$. The choice of $\ell_p$, $R_1$, and $R_2$ influences the magnetic flux $\Phi_p$ picked up by the loop as well as the self-inductance for the loop, hence the current signal $I_{\rm ind}=-\Phi_p/L$ induced in the superconducting circuit. As noted in Ref. \cite{sikivie13}, an optimization is needed to get the best result. The magnetic flux through the pickup loop is
\dis{
\Phi_p(t)=&\int^{R_2}_{R_1} B_\theta(r,t)\ell_p dr
=\ell_p ignI\sqrt{T}\int^\infty_{-\infty}\frac{d\omega}{2\pi}f(\omega){\cal A}(\omega)\\
&\cdot \frac{1}{\omega}(J_0(\omega R_1)-J_0(\omega R_2))Q_J(\omega)e^{-i\phi_J(\omega)}e^{-i\omega t}.
}
Since the self-inductance for the loop is $L_p\simeq (\ell_p/\pi)\ln ((R_2-R_1)/a)$ where $a$ is the radius of the wire, by lengthening the loop we gain only  when the $L_p$ is negligible compared to the self-inductance of the rest of the circuit, which is usually not the case.  An incremental improvement of signal through optimization would be very important when the signal strength is on a par with the experimental sensitivity .  In our case, however, still the most dramatic improvement of the signal can come from the resonance enhancement via $Q_J$, which was not exploited in the previous work \cite{sikivie13}.
When our tube geometry hits one of the resonance points, we could have the induced current signal enhanced by the large factor $Q_J\sim 10^{3-5}$. Then the signal strength falls well within the limit of experimental sensitivity of currently available best magnetometers.


\vskip 1cm
\noindent{\bf Acknowledgments}

We would like to thank Kiwoon Choi and Beomseok Kaye for helpful
discussions. This work is supported by the IBS(IBS CA1310). J.E.K. is
also supported in part by the National Research Foundation (NRF) grant
funded by the Korean Government (MEST) (No. 2005-0093841).

\end{document}